\documentclass[conference,10pt]{IEEEtran}
\IEEEoverridecommandlockouts
\usepackage{cite}
\usepackage[printonlyused,nolist]{acronym}
\usepackage{amsmath,amssymb,amsfonts,amsthm}
\usepackage{algorithmic}
\usepackage[linesnumbered,ruled,vlined]{algorithm2e}
\usepackage{caption} 
\usepackage{subcaption}
\usepackage{enumitem}
\usepackage{graphicx}
\usepackage{textcomp}
\usepackage{xcolor}
\usepackage{makecell}

\def\BibTeX{{\rm B\kern-.05em{\sc i\kern-.025em b}\kern-.08em
    T\kern-.1667em\lower.7ex\hbox{E}\kern-.125emX}}
\usepackage{multirow}
\usepackage{url}
\usepackage[many]{tcolorbox}

\def\BibTeX{{\rm B\kern-.05em{\sc i\kern-.025em b}\kern-.08em
    T\kern-.1667em\lower.7ex\hbox{E}\kern-.125emX}}
\usepackage{pifont}

\begin{document}
\setlength{\intextsep}{1pt}
\setlength{\textfloatsep}{1pt}
\setlength{\abovecaptionskip}{1pt}
\setlength{\belowcaptionskip}{1pt}
\IEEEaftertitletext{\vspace{-1\baselineskip}}

\begin{acronym}
\acro{DNN}[DNN]{Deep Neural Network}
\acro{CNN}[CNN]{Convolutional Neural Network}
\acro{AI}[AI]{Artificial Intelligence}
\acro{LLM}[LLM]{Large Language Model}
\acro{GPU}[GPU]{Graphics Processing Unit}
\acro{SM}[SM]{Streaming Multiprocessor}
\acro{HBM}[HBM]{High-Bandwidth Memory}
\acro{ML}[ML]{Machine Learning}
\acro{MLP}[MLP]{Multi-Layer Perceptron}
\end{acronym}

\title{Technology solutions targeting the performance of gen-AI inference in resource constrained platforms}


\author{\fontsize{11}{11}\selectfont Joyjit Kundu, Joshua Klein, Aakash Patel, Dwaipayan Biswas\\
\fontsize{10}{12}\selectfont Interuniversity Microelectronics Centre (IMEC)\\ Kapeldreef 75,  
3001 Leuven, Belgium \\ Email: \{joyjit.kundu, joshua.klein, aakash.patel, dwaipayan.biswas\}@imec.be}
\maketitle
\begin{abstract}
The rise of generative AI workloads, particularly language model inference, is intensifying on/off-chip memory pressure. Multimodal inputs such as video streams or images and downstream applications like Question Answering (QA) and analysis over large documents incur long context lengths, requiring caching of massive Key and Value states of the previous tokens. Even a low degree of concurrent inference serving on resource-constrained devices, like mobiles, can further add to memory capacity pressure and runtime memory management complexity. In this paper, we evaluate the performance implications of two emerging technology solutions to alleviate the memory pressure in terms of both capacity and bandwidth using a hierarchical roofline-based analytical performance model. For large models (e.g., 13B parameters) and context lengths, we investigate the performance implications of High Bandwidth Storage (HBS) and outline bandwidth/latency requirements to achieve an acceptable throughput for interactivity. For small models (e.g., 1B parameters), we evaluate the merit of a bonded global buffer memory chiplet and propose how to best utilize it.
\end{abstract}


\section{Introduction}
The advent of Large Language Models (LLMs) has been nothing short of a disruption across all industries \cite{10.1145/3641289}. On-device (e.g., smartphone) deployment of such models, is especially interesting since it paves the path towards secure intelligent personal assistants\cite{DEBARCELOSSILVA2020113193,agent-xpu}. However, typical language models with unprecedented performance often have billions, if not trillions of parameters \cite{grattafiori2024llama3herdmodels, eff_megatron}, demanding extremely high compute and memory requirements to run. For instance, a Llama-7B model with single precision (FP16) needs around 14 GB of memory just to hold the weights, exceeding the capacity of standard DDRs in a personal device, like a smartphone. 
During inference a sequence of input tokens or prefill is processed to generate the next tokens one by one. Context length is defined by the sum of prefill and generation length. Each token carries its associated Query, Key, and Value (QKV) vectors; the Keys and Values of all previous tokens are reused to compute attention for every token generation. To avoid recalculating them, models cache K and V matrices (KV cache). The KV cache takes up additional memory space and grows linearly with the context length \cite{davies2025liminalexploringfrontiersllm}. Multimodal inputs and concurrent inference serving further worsen the memory management issue by extending context length and adding runtime complexity \cite{Inf-MLLM}. 
The on-chip memory or standard SRAM is often not sufficient to hold the model and the KV cache. Fetching them from off-chip memory and generating only one token at a time results in ops/bytes $\sim \mathcal{O}(1)$. Now, due to the {\em memory wall} problem \cite{AIMemWall} data movement becomes the major bottleneck in the generation phase of inference -- this nature of the workload is termed as memory-boundedness.

Various algorithmic solutions have been devised to tackle this problem, primarily capitalizing on the concept of dynamic memory management techniques that enable larger models and longer contexts with acceptable throughput. Some of the ideas include, attention score-based KV cache eviction with large context lengths on edge platforms yielding up to 2x speedup \cite{Inf-MLLM}, smart memory management techniques like block access from Flash and sparsity-aware loading achieving 3–20x speedup depending on the baseline and model type for iPhone SoCs \cite{apple-flash}. Utilization of runtime optimizations such as concurrent inference serving case, workload-aware scheduling, disaggregated prefill \& decode, and heterogeneous-Soc aware batching can deliver an additional 1.6–4.8x improvement \cite{agent-xpu}. Such algorithmic and middleware-level improvements can always be augmented with technology solutions.

Large memory requirements of model weights and KV cache typically rely on SSD/Flash storage. However, they are orders of magnitude slower (in bandwidth) than DDR \cite{apple-flash, flex-gen}. To fill this gap, high bandwidth storage (HBS) promises bandwidth comparable with DDR with orders of magnitude more capacity at the cost of latency (microseconds) \cite{kioxia}. 
An additional augmentation to storage (with or without HBS) can be a SRAM based hybrid-bonded buffer memory chiplet that uses a custom interface directly to the logic die-resident NPU. 
This proposal draws inspiration from compute-near/in-memory works that usually assume the logic die to be close to memory \cite{H3D-Transformer}. 
In this paper, we explore the performance implications of the above two technology propositions. Our main contributions are the following:
\begin{itemize}
    \item Within an analytical framework, we show a methodology of evaluating the performance of emerging technologies for gen-AI inference and further offer insights into the key factors behind their projected performance, thus, facilitating System-Technology co-optimization (STCO).   
    \item For large models, we explore bandwidth-latency requirements of High Bandwidth Storage (HBS) memory to achieve a throughput threshold for various context windows. 
    \item For small models, we investigate the performance of a SRAM based chiplet solution for Q, K, V matrices and weights. 
\end{itemize}
\section{Methodology}
\label{method}
We adapt an existing analytical performance modeling framework validated across a number of accelerators, LLM models, and mapping choices both for training and inference to incorporate the additional memory subsystem, namely, HBS and bonded memory chiplet \cite{optimus1, optimus2}.  Most of the other available performance models typically consider roofline analysis with only one memory level \cite{calculon} and thus, often fail to capture intricacies of complex memory hierarchy. The core of the performance model used here relies on a hardware-aware {\em hierarchical roofline} analysis. It derives the compute/task-graph of the language models involving kernels like, GEMMs/GEMVs (General Matrix-Matrix or Matrix-Vector multiplications) and ones with element-wise operations. The performance is dominated by the following GEMM kernels: \{X$.{\rm W_{Q/K/V}} = {\rm Q/K/V}$\}, \{Q.${\rm K^T}$ = R\}, \{Softmax(R).V = Z\}, \{Z.${\rm W_O}$ = O\} (projection), \{O.${\rm W_{MLP1}} = {\rm O_{MLP1}}$\}, \{${\rm O_{MLP1}}$.${\rm W_{MLP2}} = {\rm O_{MLP2}}$\}, where W stands for the weight matrices and X for input. 
Respecting sequential dependencies, each kernel is mapped onto the given system architecture. GEMM/GEMV kernels are parallelized via tiling distributed across the processing elements, with tile sizes determined by cache and memory capacities. The memory access pattern of the above kernels are quite regular and thus, predictable analytically. Kernel latency is estimated by searching over candidate tiling strategies at each memory hierarchy. For a given level, the optimal tiling defines the memory access count, from which the arithmetic intensity (AI)—the ratio of FLOPs to bytes transferred—is computed. From the hardware perspective, each memory level has an inflection point defined by the ratio of the compute throughput to the memory bandwidth. 
If a kernel’s AI is below this point, it is memory-bound; otherwise, it is compute-bound. The kernel execution time is given by the critical path across all memory levels:
\begin{equation}
{\rm time} = max \{\frac{{\rm total~FLOPs}}{{\rm compute~throughput}}, \frac{{\rm data~traffic~volume}}{\rm memory~throughput}\}
\end{equation}
We consider the following base memory hierarchy: NPU scratchpad – L2 – DDR where each memory level is characterized by its capacity, latency, and bandwidth \cite{DeepFlow}. 
\section{Experiment Design \& Results}
\begin{figure}
\centering
\includegraphics[width=\linewidth]{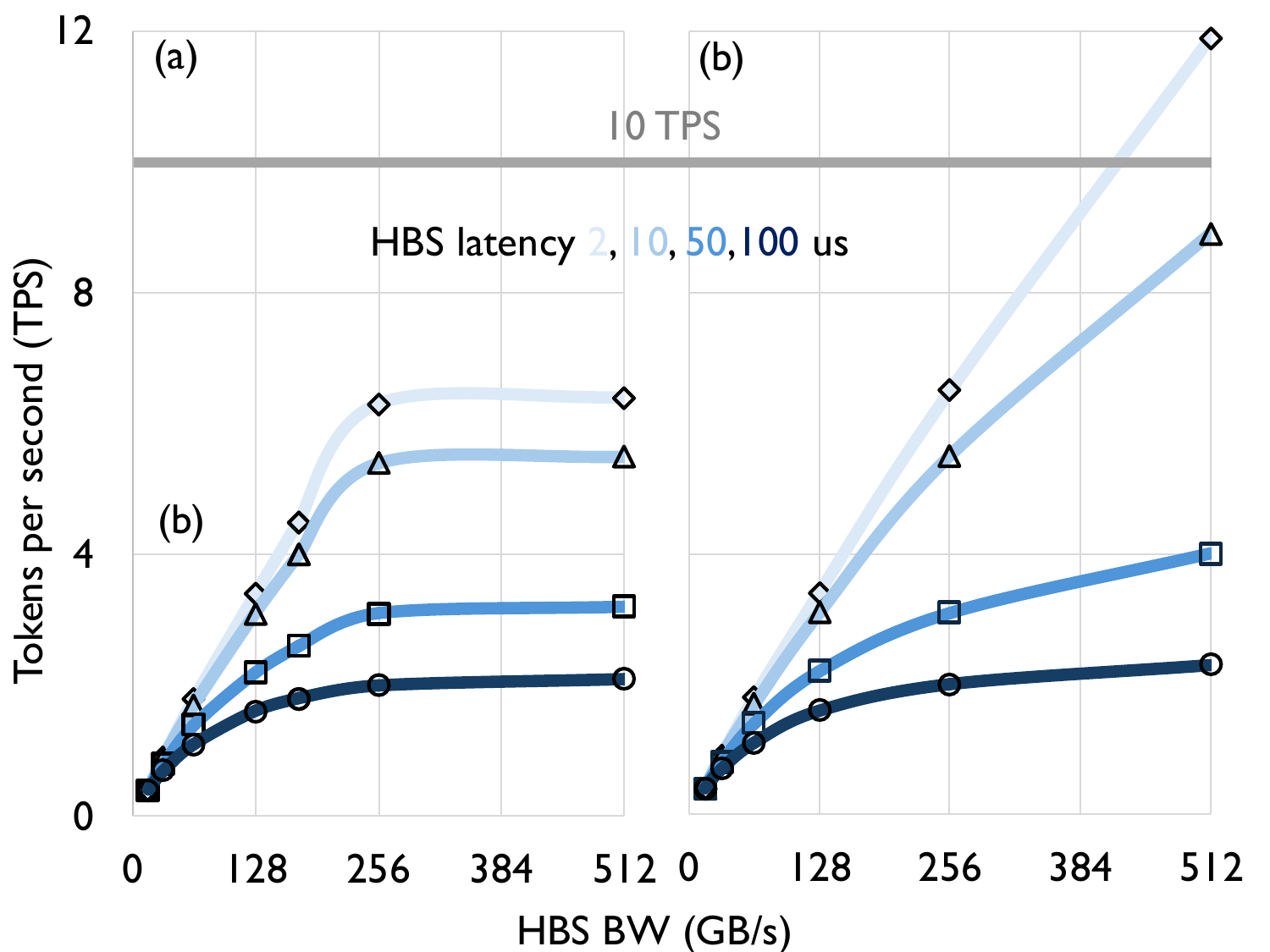}
\caption{TPS as a function of HBS bandwidth and latency when DDR bandwidth is set at (a) 173 GB/s (LPDDR6), (b) 520 GB/s (3x stacked LPDDR6). The data are for LLaVa1.5-13B with single precision. DDR latency is fixed at 100 ns.}
\label{fig:fig02}
\end{figure}


\begin{figure}
\centering
\includegraphics[width=\linewidth]{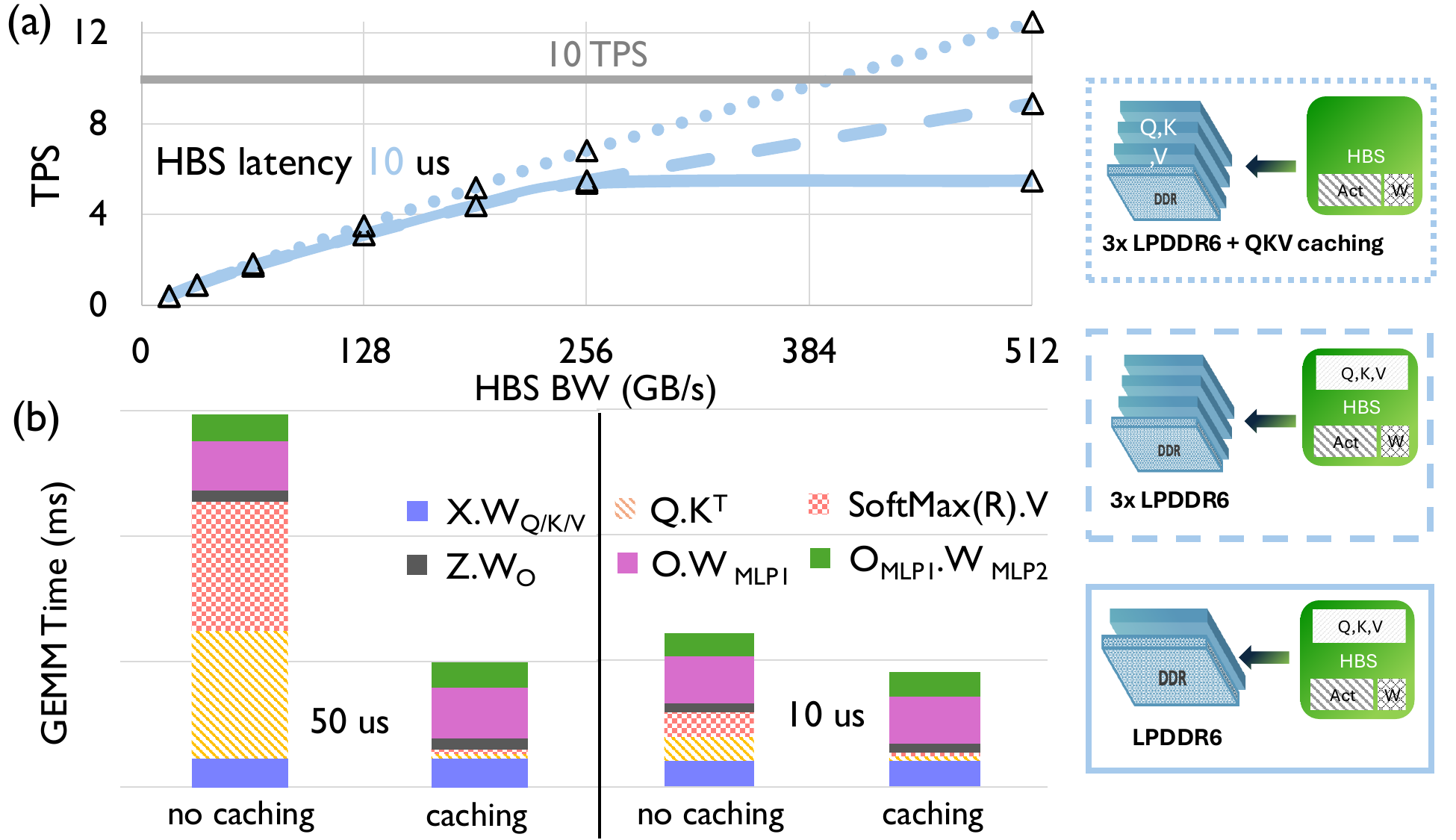}
\caption{(a) TPS as a function of HBS bandwidth when latency is fixed at 10 $\mu$s for three different configurations of DDR: with I)  LPDDR6, II) 3x LPDDR6 -- in both of these cases Q, K, V, activations and weights are in HBS, III) 3x LPDDR6 and the attention traffic is restricted to DDR. (b) Time (per layer) spent in different GEMM operations (see Sec.~\ref{method}) for two different HBS latencies while bandwidth is fixed at 512 GB/s. The ones involved in self-attention are patterned.}
\label{fig:fig04}
\end{figure}


First, we set up the experiments for HBS. We choose the multimodal language model LLaVa1.5-13B at single precision with three sets of prefill/decode lengths: 200/200, 4096/12288, and 8192/24576. The inputs are image and text, both expressed in terms of a sequence of tokens and being concatenated for processing through the model -- thus, multimodality entails a bigger context length. We consider a single NPU compute instance in isolation with a total compute throughput of 35 TFLOPs  (combining all the processing elements) and the following memory hierarchy: NPU scratchpad -- L2 -- DDR -- HBS. HBS capacity is considered to be big enough to hold the model and the required Q matrix, KV cache, and the intermediate activations.

As offloading storage solutions we consider current Flash-based SSDs with PCIe Gen5 and Gen6 interfaces allowing 16 and 32 GBps of throughput. For scaling further, we consider HBS containing 16 IO per plane with 1-4 Gbps of per IO data-rate with interleaving, overall throughput is then considered based on variations of plane and rank structure. Such high data-rates are achieved by having much more smaller planes compared to conventional SSDs. Thus, HBS has potential to offer order of magnitude more bandwidth compared to traditional SSDs sacrificing SSD like density/capacity. 


We set the LPDDR6 latency at 100 ns otherwise specified and design the following three experiments sweeping HBS bandwidth and latency to measure throughput in terms of tokens per second (TPS). We set a target of 10 TPS that is minimum to achieve any interactivity. 
\begin{itemize}
    \item[I.] DDR bandwidth is fixed at 173 GB/s (LPDDR6). HBS bandwidth is varied from 16 to 512 GB/s for four latency values: 2, 10, 50 and 100 us.
    \item[II.] The DDR bandwidth is set at 3x of LPDDR6 ($\sim$ 520 GB/s) while HBS bandwidth is varied from 16-512 GB/s. For the above cases, Q, K, V, weights and activations, by default, reside on HBS and hierarchical tiling is used to bring them towards on-chip registers. 
    \item[III.] Q, K, V matrices and the intermediate activations are selectively placed in DDR to reduce traffic between HBS and DDR. The remaining set up is same as the above. 
\end{itemize}


\begin{table}[h!]
\caption{Performance of different DDR (latency 100 ns) and HBS set ups without and with restricted DDR-HBS traffic.}
\label{fig:fig07}
\centering
 \begin{tabular}{c c c c} 
 \hline
 \textbf{DDR BW} & \textbf{HBS BW} & \textbf{\makecell{Q, K, V, \\ Act storage}} & \textbf{Performance} \\
 \hline \hline
 \makecell{LPDDR6 \\ (173 GB/s)} & 16-173 GB/s & HBS & \makecell{TPS $\sim$ 4 \\ bottleneck- HBS} \\ \hline
 \makecell{LPDDR6 \\ (173 GB/s)} & 16-520 GB/s & HBS & \makecell{TPS $\sim$ 5.5 \\ Bottleneck- DDR \\ $1.4\times$ gain}\\ \hline 
 \makecell{3x stacked \\ LPDDR6\\ (520 GB/s)} & 16-520 GB/s & HBS & \makecell{TPS $\sim$ 8.9 \\ bottleneck- HBS \\ $2.2\times$ gain} \\ \hline
 \makecell{3x stacked \\ LPDDR6\\ (520 GB/s)} & 16-520 GB/s & \makecell{Q,K,V in DDR\\
rest in HBS} & \makecell{TPS $\sim$ 12.5 \\ bottleneck- HBS \\ $3.1\times$ gain} \\
 \hline
\label{fig:fig07}
 \end{tabular}
\end{table}

The results of Experiment I are shown in Fig.~\ref{fig:fig02}(a). DDR bandwidth is fixed at 173 GB/s (LPDDR6), while HBS bandwidth is swept across different latency values. When HBS bandwidth is below DDR bandwidth, HBS is the performance bottleneck and thus, tokens per second (TPS) scales linearly with HBS bandwidth, with higher HBS latency yielding lower throughput. Even when the HBS bandwidth slightly exceeds the DDR bandwidth, its much higher latency can keep HBS as the bottleneck, so, TPS may continue to scale depending on the data-movement requirements of the kernel. When HBS bandwidth is $\gtrsim 40\%$ of LPDDR6 bandwidth, the bottleneck shifts to DDR, causing TPS to saturate well below 10 TPS. In Experiment II (Fig.~\ref{fig:fig02}(b)), increasing DDR bandwidth by 3× (e.g., via wider I/O or 3D stacking) removes the DDR bottleneck, shifting it back to HBS and allowing TPS to continue scaling with HBS bandwidth. Only the 2 $\mu$s HBS-latency curve exceeds the 10 TPS target only at very high HBS bandwidths. In both experiments, data reside primarily in the HBS and are tiled across memory levels to maximize reuse.

{\em \textbf{Key takeaway I.} HBS remains the performance bottleneck even when its bandwidth is slightly above the DDR bandwidth due to its high latency; the bottleneck shifts when HBS bandwidth is at least, $40\%$ higher than that of DDR.}

\begin{figure}
\centering
\includegraphics[width=\linewidth]{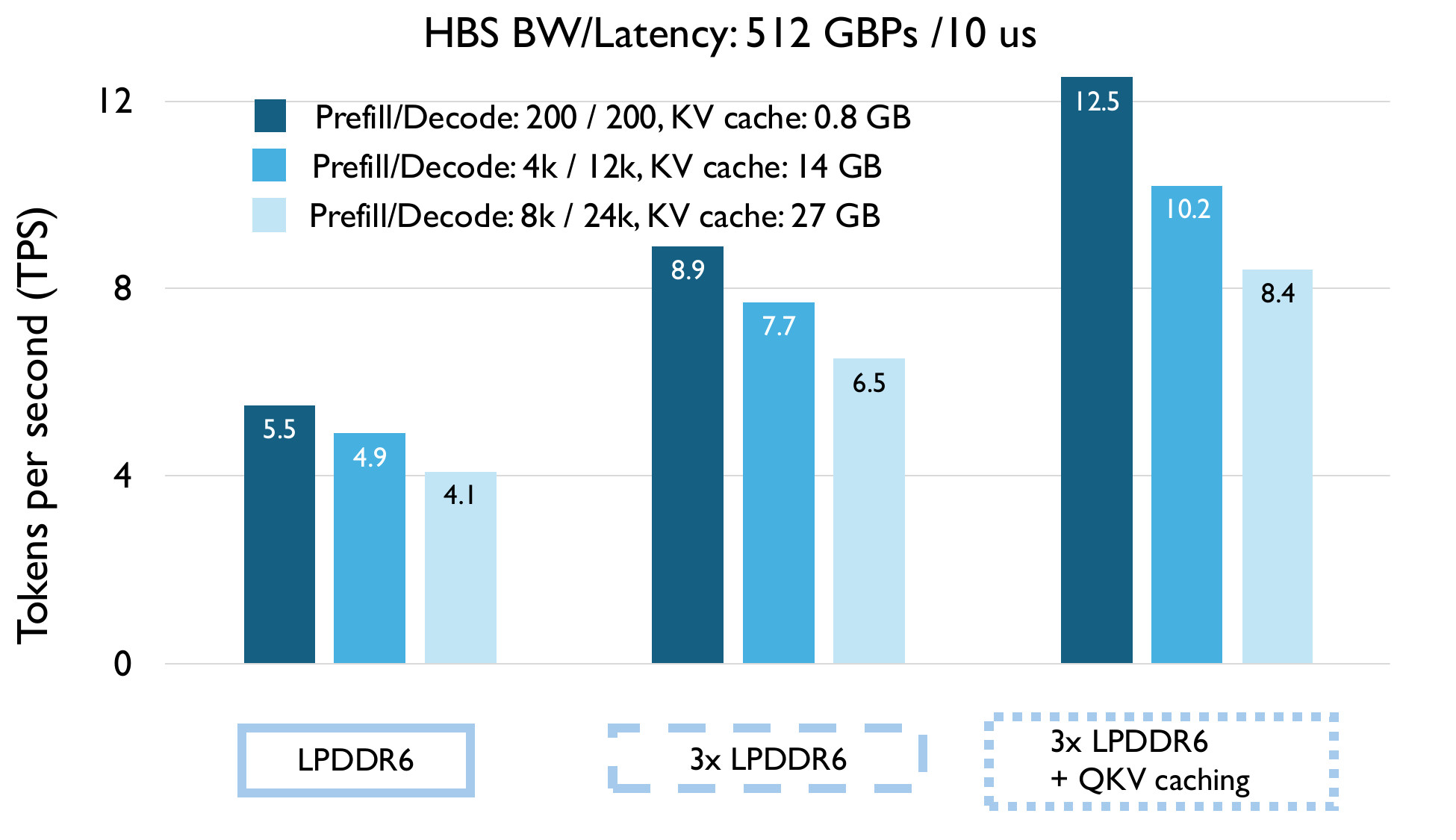}
\caption{Achieved TPS for the three different configurations of DDR: bandwidth set at I) LPDDR6, II) 3x LPDDR6, III) 3x LPDDR6 with Q, K, V cached at DDR. HBS bandwidth and latency are fixed at 512 GB/s and 10 us. The data are for three different context windows.}
\label{fig:fig05}
\end{figure}

In Experiment III, data movement related to Q, K, and V is restricted to DDR, reducing traffic between DDR and HBS. To combat SRAM's scaling limitations, DDR can be treated as a last-level cache by enabling small, low-latency mats to store tags and metadata, and opportunistic tag checks \cite{TDRAM}. Figure~\ref{fig:fig04}(a) compares all three scenarios; results are shown for 10 $\mu$s HBS latency. Limiting Q/K/V related data transfers improves attention execution time which accounts for 31–69\% of the total GEMM time depending on HBS latency—and enables achieving the 10 TPS target even at 10 $\mu$s latency, as shown in Fig.~\ref{fig:fig04}. A summary is provided in Table~\ref{fig:fig07}. Using the configuration with DDR bandwidth fixed at 173 GB/s and HBS bandwidth swept from 16 GB/s up to DDR bandwidth at 10 $\mu$s latency as the baseline, Table~\ref{fig:fig07} reports the performance impact of the three experiments. With 3$\times$ LPDDR6 bandwidth, 512 GB/s HBS bandwidth, and Q/K/V cached in DDR, a 3.1$\times$ speedup is achieved, corresponding to $\sim$12.5 TPS at 10 $\mu$s HBS latency. All results are for a prefill/decode length of 200/200. Similar trends hold for larger context lengths as shown in Figure~\ref{fig:fig05} where we vary prefill/decode up to 8192/24576. The required KV-cache size for a 33k context is $\sim$27 GB. For all configurations, performance degrades monotonically with increasing context length, as each generated token contextualize to all previous tokens. However, the relative performance gains across the three experiments remain consistent across all context lengths.

{\em \textbf{Key takeaway II.} 3D stacking or wider-IO enables DDR bandwidth scaling and thus, offers more room for performance improvement with faster HBS. Still, the basic tiling approach does not attain the target performance.}

{\em \textbf{Key takeaway III.}Reducing DDR–HBS traffic yields additional speedups, especially for larger models (e.g., LLaVA-13B), where attention GEMMs account for 31–69\% of total GEMM time for HBS latencies of 10–50 $\mu$s.}


\begin{figure}
\centering
\includegraphics[width=\linewidth]{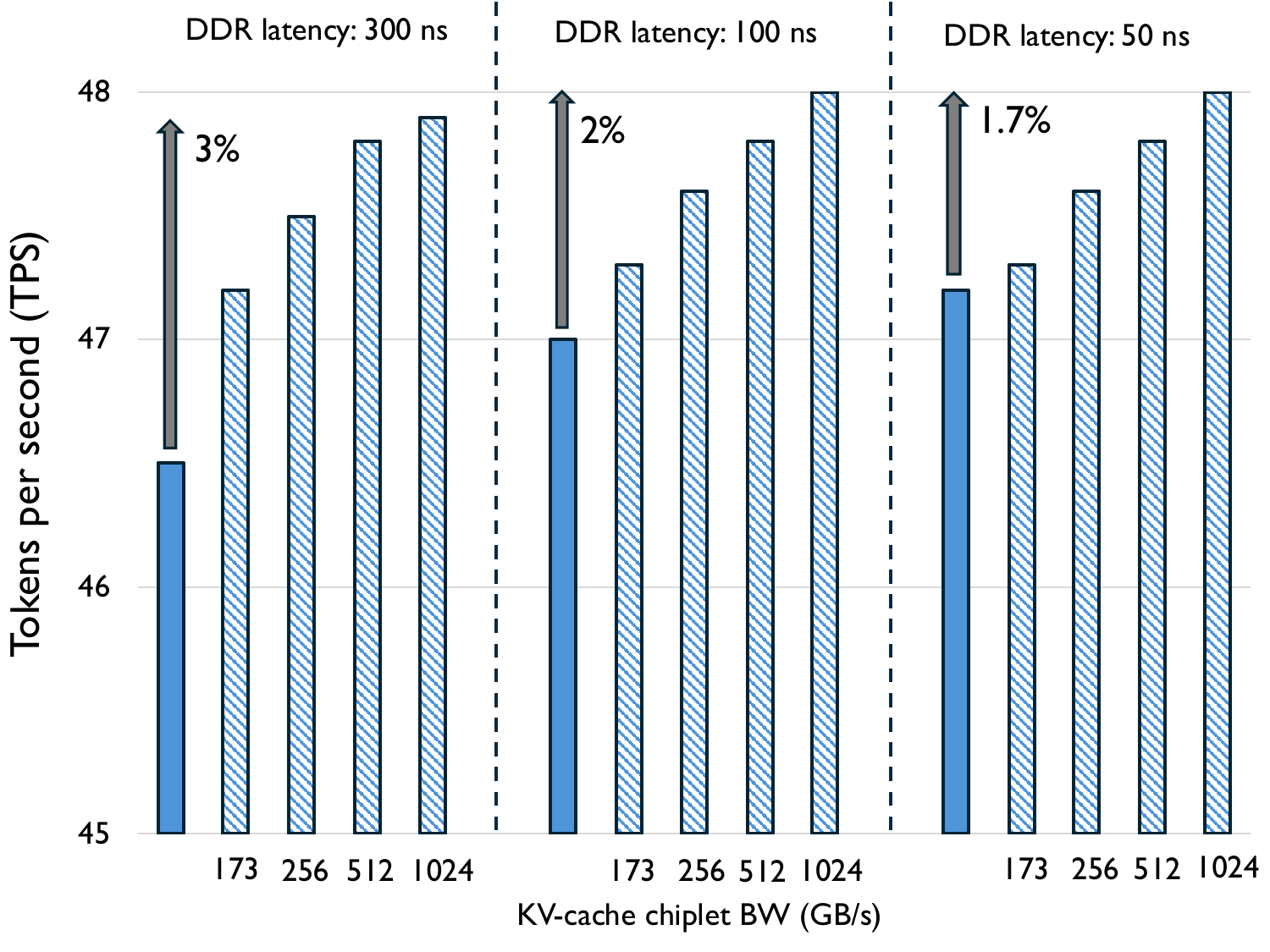}
\caption{Performance implications of QKV chiplet for Llama3.2-1B model while the bandwidth of the chiplet is varied for different latency configurations of DDR with the bandwidth fixed at 173 GB/s. The solid bars indicate the baseline performance with no chiplet configuration.}
\label{fig:fig09}
\end{figure}


Next, we investigate the potential of a SRAM based global chiplet buffer for inference with smaller models -- as a test case, we consider Llama3.2-1B with prefill/decode length 128/384. The modeled memory subsystem is as follows: NPU scratchpad -- L2 + chiplet for QKV -- DDR. As NPUs are largely orchestrated ASICs with fine-grained control of memory (i.e., they bypass virtual memory in favor of direct memory accesses), we envision that Q and KV cache accesses could be tagged such that they are redirected to the bonded chiplet, connected via custom interface. From a software perspective, such a memory could sit next to NPU L2 cache weights sit in L2 and Q, K, V traffic goes to the chiplet. 

The required KV cache size is $\sim 68 $ MB. Since the chiplet is SRAM based, we consider a size ($> 68$ MB) that is enough to hold the Q and KV-cache of smaller models. Note that the chiplet is placed at the same footing as L2. For different DDR latencies, we sweep the chiplet bandwidth from 173 GB/s to 1 TB/s. The throughput results are shown in Fig.~\ref{fig:fig09}. The solid bars are for the cases in the absence of the bonded memory chiplet. All the GEMM kernels during decode are DDR bandwith bound. With introducing the bonded chiplet, we selectively restrict the data movement related to Q, K, V, and the intermediate activations during attention computation to the the chiplet, resulting in faster attention computation. 
If the DDR latency is higher, the performance gain is enhanced (see Fig.~\ref{fig:fig09}). Note that the degree of improvement is not as high as we have seen in case of the HBS studies. Reason being for small models, the GEMMs involved in the attention computation of the decode phase amount only to 4-9\% of the total GEMM time. While, the projection and MLP part account for 82-86\%. Thus, a better strategy for acceleration would be to use the chiplet for related weights or activations. 


{\em \textbf{Key takeaway IV.} While a chiplet for Q, K, V accelerates attention, its benefit is limited for smaller models (e.g., LLaMA-3.2-1B), where attention GEMMs account for only 4–9\% of total GEMM time (for DDR latency 0.1–1 $\mu$s). Greater performance gains can be achieved by using it for MLP and projection weight matrices instead.}

\section{Conclusion}
In summary, we investigated the performance implications of two technological propositions: High bandwidth storage for larger models and an SRAM-based bonded chiplet for smaller models to alleviate the memory pressure of on-device gen-AI inference. We estimate the latency and bandwidth requirements for HBS to achieve a throughput of 10 tokens per second for a multimodal model, LLaVa1.5-13B. For smaller models, we explore the potential of a chiplet solution with SRAM technology and what is the most performant use of that. By exposing the time breakdown of different GEMM operations within the execution graph, we show that for large models, caching Q, K, and V is efficient but for smaller models, it has limited performance gain. Caching weight matrices for MLP and projection computation is a more performant solution. These solutions can potentially relax the memory pressure of gen AI workloads without much algorithmic intervention. 

\bibliographystyle{IEEEtran}

\end{document}